\documentclass[aps,pre,reprint,groupedaddress,amsmath,amssymb,showpacs]{revtex4-1}
\usepackage{graphicx}
\usepackage{CJK}
\begin{document}
\begin{CJK*}{UTF8}{}
\title{Deformation of an asymmetric thin film}
\author{Jun Geng (\CJKfamily{gbsn}耿君)}
\author{Jonathan V. Selinger}
\email{jselinge@kent.edu}
\affiliation{Liquid Crystal Institute, Kent State University, Kent, Ohio 44242, USA}
\date{November 2, 2011}
\begin{abstract}
Experiments have investigated shape changes of polymer films induced by asymmetric swelling by a chemical vapor.  Inspired by recent work on the shaping of elastic sheets by non-Euclidean metrics [Y.~Klein, E.~Efrati, and E.~Sharon, Science \textbf{315}, 1116 (2007)], we represent the effect of chemical vapors by a change in the target metric tensor.  In this problem, unlike that earlier work, the target metric is asymmetric between the two sides of the film.  Changing this metric induces a curvature of the film, which may be curvature into a partial cylinder or a partial sphere.  We calculate the elastic energy for each of these shapes, and show that the sphere is favored for films smaller than a critical size, which depends on the film thickness, while the cylinder is favored for larger films.
\end{abstract}
\maketitle
\end{CJK*}

\section{Introduction}

Thin films are three dimensional (3D) objects with one dimension much smaller than the other two.  Such films are not always flat; they can easily form buckled or wrinkled 3D shapes.  Over several years, there has been extensive theoretical and experimental research to explore the mechanisms for shape selection.  This research is important to understand the formation of biological structures, such as leaves and flowers, in which thin films assume well-defined shapes with biological functions.  It is also important for the design of new synthetic materials, which should spontaneously form desired morphologies~\cite{Finot1996,Freund1996,Freund2000,Long2010}.

Sharon and collaborators have recently developed an important theoretical approach for addressing shape selection in thin elastic sheets~\cite{ Klein2007,Efrati2007,Efrati2009,Efrati2009a,Sharon2010,Kamien2007}.  In this approach, a film is characterized by a ``target metric tensor,'' which describes the ideal spacings between points in the film that minimize the local energy.  Depending on the mathematical properties of this tensor, there may or may not be any global shape of the film embedded in 3D Euclidean space that achieves the ideal spacings everywhere.  If this state is not achievable, the film is geometrically frustrated.  Its lowest-energy state will then have residual local stresses and strains, and will generally be curved in a complex way~\cite{Hoger1985,BenAmar2005}.  Sharon \emph{et al.} have demonstrated this approach experimentally by using thin films of gels, which can be expanded locally by adding a nonuniform concentration of a dopant.  The gels then relax to the 3D shape that has been programmed by the dopant concentration profile, in agreement with the geometric calculations.

One limitation of Sharon's theory is that it assumes the films are uniform across their thickness.  This limitation is significant, because many types of films have some variation in their elastic properties across their thickness. Indeed, this type of variation might provide an additional way to design films to form desired structures.  The purpose of this paper is to generalize the theory to describe asymmetric films, with arbitrary variation across the thickness.  Through this generalization, we show that the asymmetry leads to new types of terms in the elastic free energy of the film, and we investigate how these terms change the curvature.

\begin{figure}[b]
\includegraphics[width=3in]{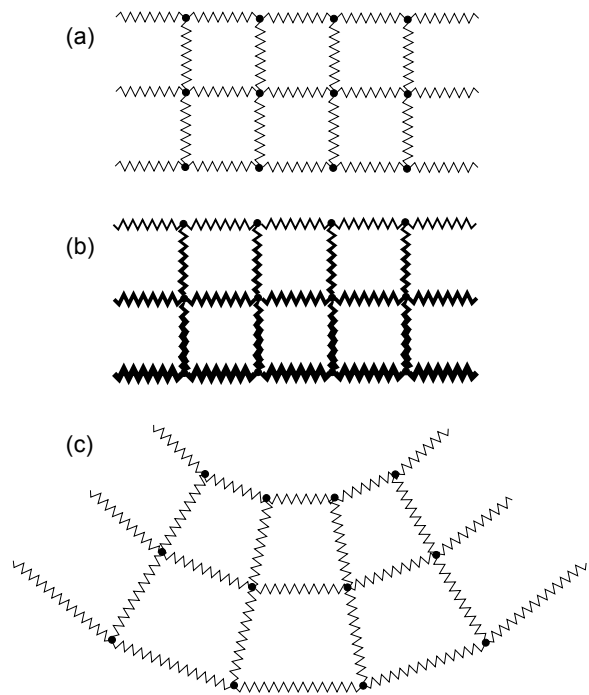}
\caption{\label{fig:shell}Schematic illustration of the deformation of a thin shell due to swelling: (a)~before swelling, (b)~after swelling without any deformation, and (c)~deformed shell.}
\end{figure}

As a specific example to motivate this study, we consider the asymmetric swelling of a thin film by absorption of a gas or liquid.  A schematic view of this problem is shown in Fig.~\ref{fig:shell}.  Here, a cross section through the film is illustrated by mass points connected by springs, with bold lines indicating that a spring is compressed relative to its stress-free length. Before swelling, in Fig.~\ref{fig:shell}(a), all the springs are at the stress-free length, and the film is flat.  After swelling, the intrinsic stress-free lengths of the springs vary gradually along the thickness direction, as shown in Fig.~\ref{fig:shell}(b).  Hence, the film will deform to minimize the energy, as shown in Fig.~\ref{fig:shell}(c).  Because the film is 3D, the actual geometry is more complex than the cross section shown in the figure.  It is not obvious whether the film should deform into a partial cylinder (with mean curvature but no Gaussian curvature) or a partial sphere (with both mean and Gaussian curvature).  In either case, some of the springs will be unable to achieve their intrinsic stress-free length.  Hence, this is a simple example of a geometrically frustrated structure.

To address this problem, we extend the approach of Sharon \emph{et al.} to consider a target metric that depends on position across the thickness of the film.  We calculate the energy of the deformed film in cylindrical and spherical geometries, and find there is a critical lateral size, which depends on the film thickness and the gradient of the intrinsic metric.  If the lateral size of the film is smaller than this critical size, a partial sphere is preferred; otherwise, a partial cylinder has lower energy.

This paper is organized as follows.  In Sec.~\ref{sec:theory}, we briefly review the theory of non-Euclidean plates, and use it to calculate the energy in the general case where the target metric is proportional to the distance from the midplane.  We then use this general formula to calculate the energy for a cylinder in Sec.~\ref{sec:cylinder} and for a sphere in Sec.~\ref{sec:sphere}.  Finally, in Sec.~\ref{sec:discuss}, we discuss the results of this study and compare with previous work on related elastic problems.

\section{\label{sec:theory}Theory}

In the first part of this section, we briefly review the non-Euclidean theory developed by Sharon~\emph{et al.} for the special case in which the target metric is uniform across the thickness of the film~\cite{Efrati2009, Efrati2009a}.  We then introduce our calculation for the asymmetric film with a target metric tensor that depends on the distance from the midplane of the film.

\subsection{Brief review of non-Euclidean theory}

Following Sharon~\emph{et al.}, we define the elastic free energy for any configuration of a material through the following construction. First, we construct a coordinate system $x^i$, for $i=1,2,3$, in the material frame.  The current 3D position of a point $(x^1,x^2,x^3)$ is then given by $\mathbf{R}(x^1,x^2,x^3)$.  As a result, the current spacing between two nearby material points $(x^1,x^2,x^3)$ and $(x^1+dx^1,x^2+dx^2,x^3+dx^3)$ is given by $(ds)^2 = g_{ij} dx^i dx^j$ (using the Einstein summation convention, where Latin indices range from 1 to 3, and Greek indices range from 1 to 2).  Here, the metric tensor is $g_{ij}=\partial_{x^i}\mathbf{R}\cdot\partial_{x^j}\mathbf{R} =\partial_i\mathbf{R}\cdot\partial_j\mathbf{R}$.  By comparison, if we were to cut out a very small piece of material around the point $(x^1,x^2,x^3)$, and let this small piece relax to its lowest-energy state, then the spacing between those points would be $(\overline{ds})^2 = \bar{g}_{ij} dx^i dx^j$, where $\bar{g}_{ij}$ is the target metric tensor that characterizes the intrinsic, ideal spacing between the points.  The Green-Lagrangian strain tensor is then defined as the difference between the current metric and the target metric
\begin{equation}
\varepsilon_{ij}=\frac{1}{2}\left(g_{ij}-\bar{g}_{ij}\right).
\label{eq:strain}
\end{equation}
If the material is isotropic, the most general expression for the 3D elastic energy density can be written as
\begin{equation}
\label{eq:energydensity3d}
f_3 = \frac{1}{2}A^{ijkl}\varepsilon _{ij}\varepsilon_{kl},
\end{equation}
where
\begin{equation}
A^{ijkl}=\lambda\bar{g}^{ij}\bar{g}^{kl}
+\mu(\bar{g}^{ik}\bar{g}^{jl}+\bar{g}^{il}\bar{g}^{jk}),
\end{equation}
are the contravariant components of the three-dimensional elasticity tensor in curvilinear coordinates~\cite{Ciarlet2005}.  If the target metric $\bar{g}_{ij}$ is independent of the coordinate $x^3$ across the thickness of the cell, then the 3D elastic energy density can be integrated across thickness to construct the 2D elastic energy density of the plate,
\begin{equation}
f_2 = \int f_3 dx^3.
\end{equation}
This 2D elastic energy density includes a stretching energy (proportional to thickness $h$) and a bending energy (proportional to $h^3$).  The total elastic energy then becomes
\begin{equation}
F=\int f_2 \sqrt{|\bar{g}|} dx^1 dx^2
=\int f_3 \sqrt{|\bar{g}|} dx^1 dx^2 dx^3,
\end{equation}
integrated over the whole body.

Note that the target metric tensor $\bar{g}_{ij}$ represents an ideal \emph{local} configuration of the plate with zero stress and zero elastic energy density.  There might or might not be any \emph{global} configuration in 3D Euclidean space that has this target metric tensor everywhere.  \emph{If such a configuration exists}, then it is a stress-free reference configuration, which can be used to formulate Truesdell's hyper-elasticity principle~\cite{Truesdell1966}.  On the other hand, \emph{if such a configuration does not exist}, then there is no such stress-free reference configuration.  It is still possible to minimize the elastic energy to find the equilibrium state of the plate, but this equilibrium state must be a frustrated state with residual local stress.  In this case, the object can be called a non-Euclidean plane, whose midplane has no immersion with zero stretching in 3D Euclidean space.

\subsection{Asymmetric film}

In our problem of asymmetric swelling of a thin film, as shown in Fig.~\ref{fig:shell}, the target metric must depend on position across the thickness of the film.  If the gas vapor  induces expansion of the polymer film, the expansions at the top and the bottom of the film are different due to the different concentrations of the gas.  Assuming the gas concentration varies linearly across the thickness, the target tangent vectors can be written as
\begin{equation}
\label{eq:ttangent}
\bar{\mathbf{g}}_i=\mathring{\mathbf{g}}_i\left(p+qx^{3}\right),
\end{equation}
and hence the target metric is
\begin{equation}
\label{eq:tmetric}
\bar{g}_{ij}=\bar{\mathbf{g}}_i \cdot\bar{\mathbf{g}}_j=\mathring{g}_{ij} \left(p^2+2pqx^3+q^2\left(x^3\right)^2 \right).
\end{equation}
In our notation, all symbols with rings represent quantities \emph{before swelling}; in particular, $\mathring{\mathbf{g}}_i$ are the tangent vectors and $\mathring{g}_{ij}$ is the metric before swelling.  Likewise, all symbols with bars represent target quantities \emph{after swelling}.  The coefficients $p$ and $q$ define how the gas concentration is distributed; $p$ represents a uniform expansion or compression, and $q$ represents a linear gradient in the expansion factor across the thickness of the film.

From Eq.~(\ref{eq:strain}), the strain tensor is defined as the difference between the actual metric $g_{ij}$ and the target metric $\bar{g}_{ij}$ after swelling.  To find the actual metric, we must consider a specific configuration of the plate.  Following the second Kirchhoff-Love assumption, we assume that points located on any normal to the midplane in the initial state remain on that normal in the deformed state, but the distance to the midplane may change.  Thus, a point $(x^1,x^2,x^3=0)$ \emph{on the midplane} before swelling becomes $\mathbf{R}(x^1,x^2,x^3=0)=\mathbf{R}_{\mathrm{mid}}(x^1, x^2)$ after swelling, and a point $(x^1,x^2,x^3)$ \emph{off the midplane} becomes
\begin{equation}
\label{eq:R}
\mathbf{R}(x^1, x^2, x^3)=\mathbf{R}_{\mathrm{mid}}(x^1, x^2)+\xi(x^3) \mathbf{\hat{N}}(x^1, x^2).
\end{equation}
Here, $\mathbf{\hat{N}}$ is the unit normal vector to the midplane, and $\xi(x^3)$ is the new distance to the midplane along the normal.  Because of swelling, $\xi(x^3)$ is no longer equal to $x^3$. To lowest order for a thin film, $\xi(x^3)$ can be written as power series
\begin{equation}
\label{eq:xi}
\xi(x^3)=mx^3+\frac{1}{2} m'\left(x^3\right)^2.
\end{equation}
If we assume that the local volume everywhere remains constant during deformation, equal to the swelled local volume of the target metric, then the coefficients in this expansion are constrained to be $m=p$ and $m'=3q$.

Using this configuration of the plate, we can calculate the metric tensor $g_{ij}$ as a power series in $x^3$.  From Eqs.~(\ref{eq:R}) and~(\ref{eq:xi}), the first $2\times2$ components of $g_{ij}$ can be written as
\begin{equation}
g_{\alpha \beta}=a_{\alpha \beta}-2mb_{\alpha \beta}x^3-m'b_{ \alpha \beta} \left(x^3 \right)^2,
\end{equation}
where $a_{\alpha\beta}=\partial_{\alpha}\mathbf{R}\cdot \partial_{\beta}\mathbf{R}$ is the first fundamental form of the midplane (i.~e.\ the 2D metric tensor), and $b_{\alpha\beta}={\hat{\mathbf{N}}\cdot\partial_{\alpha}\partial_{\beta}\mathbf{R}}$ is the second fundamental form of the midplane (i.~e.\ the 2D curvature tensor).  The first $2\times2$ compenents of the strain tensor are then $\varepsilon_{\alpha \beta}=\frac{1}{2}\left(g_{\alpha \beta}-\bar{g}_{\alpha \beta}\right)$.

We can now calculate the elastic energy density of the plate.  Based on the first Kirchhoff-Love assumption that the stress is in the local midplane~\cite{Love1888, Efrati2009}, we can express the full 3D elastic energy density of Eq.~(\ref{eq:energydensity3d}) in terms of the first $2\times2$ compenents of the strain tensor as
\begin{equation}
f_3=\frac{1}{2}\mathcal{A}^{\alpha \beta\gamma\delta}\varepsilon _{\alpha \beta}\varepsilon_{\gamma\delta},
\end{equation}
where
\begin{equation}
\mathcal{A}^{\alpha\beta\gamma\delta}
=2\mu\left( \frac{\lambda}{\lambda+2 \mu}\bar{g}^{\alpha\beta}\bar{g}^{\gamma\delta}
+\bar{ g}^{\alpha\gamma}\bar {g}^{\beta\delta}\right).
\end{equation}
If we assume that the local volume everywhere remains constant during deformation, then $\bar{g}^{ij}\varepsilon_{ij}=0$, and hence the elastic modulus $\lambda \to \infty$.  As a result, the elasticity tensor becomes
\begin{equation}
\mathcal{A}^{\alpha\beta\gamma\delta}=2\mu\left(\bar{g}^{\alpha\beta}\bar{ g}^ {\gamma\delta}
+\bar{g}^{\alpha\gamma}\bar{g}^{\beta\delta}\right).
\end{equation}
Note that the contravariant components of the target metric tensor in this expression are the inverse of the covariant components of Eq.~(\ref{eq:tmetric}).  Hence, the elasticity tensor depends on $x^3$ as
\begin{equation}
\mathcal{A}^{\alpha\beta\gamma\delta}=\frac{1}{p^4}\mathring{\mathcal{A}}^ { \alpha\beta\gamma\delta}
\left(1-\frac{4q}{p}x^3+\frac{10q^2}{p^2}\left(x^ 3 \right)^2 \right).
\end{equation}
Thus, the 2D elastic energy density can be calculated by integrating the 3D elastic energy density over the film thickness $w$, and neglecting terms that has higher order than $q^2$, to obtain
\begin{align}
\label{eq:f2}
f_2=&\int_{-w/2}^{w/2}f_3\mathrm{d}x^3 \nonumber  \\
=&\frac{1}{p^4}\mathring{\mathcal{A}}^{\alpha\beta\gamma\delta}
\bigg(\frac{w}{ 2}\varepsilon_{\alpha\beta}^{\mathrm{2D}}\varepsilon_{\gamma \delta}^{\mathrm{ 2D}}
+\frac{p^2w^3}{24}b_{\alpha\beta}b_{\gamma\delta} \nonumber \\
&+\frac{qw^3}{24}\varepsilon^{\mathrm{2D}}_{\alpha\beta}b_{ \gamma \delta}
+\frac{qw^3}{12}a_{\alpha\beta}b_{\gamma\delta}\nonumber \\
&+\frac{q^ 2w^3}{p^2}\varepsilon^{\mathrm{2D}}_{\alpha\beta}a_{\gamma\delta}
+ \frac{q^2w^3}{3p^2}a_{\alpha\beta}a_{\gamma\delta}\bigg).
\end{align}

To interpret this 2D elastic energy density $f_2$, note that it depends on three tensors characterizing the local geometry of the midplane:  the 2D strain tensor
\begin{equation}
\label{eq:strain2D}
\varepsilon_{\alpha\beta}^{\mathrm{2D}}=\frac{1}{2}\left( a_{\alpha\beta}-\bar{ a}_{\alpha\beta}\right),
\end{equation}
which gives the difference between the actual metric and the target metric on the midplane, as well as the 2D metric tensor $a_{\alpha\beta}$ and the 2D curvature tensor $b_{\alpha\beta}$.  If there is no swelling, so that $p=1$ and $q=0$, then $f_2$ becomes the deformation energy of a \emph{symmetric} film.  In that case, the first and second terms in Eq.~(\ref{eq:f2}) are the stretching and bending terms, and the other terms vanish.  However, if the film is swollen asymmetrically, with $q\not=0$, then the last four terms provide new couplings that are permitted by the asymmetry.  Two of these terms are \emph{odd} in the curvature tensor $b_{\alpha\beta}$, so they favor \emph{spontaneous curvature} of the film, which is then coupled to the in-plane strain $\varepsilon_{\alpha\beta}^{\mathrm{2D}}$.  The last two terms might seem to be higher order in $q$ than the others, but we will show later that the equilibrium curvature $b_{\alpha\beta}$ is proportional to $q$, and hence the last five terms are all of order $q^2$.

\begin{figure}
    \centering
\includegraphics[width=3in]{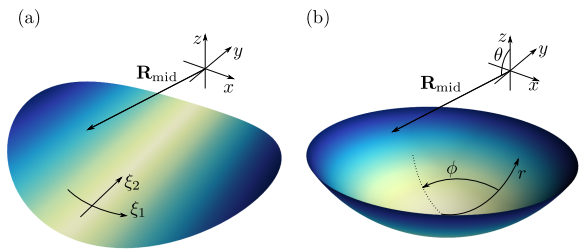}
\caption{\label{fig:shapes}(Color online) Illustration of the curved film shapes considered in Sections~\ref{sec:cylinder} and~\ref{sec:sphere}.  (a)~Partial cylinder.  (b)~Partial sphere.}
\end{figure}

For further insight into the 2D elastic energy of Eq.~(\ref{eq:f2}), in the following two sections we use it to calculate the energy of an asymmetrically swollen film curving into a partial cylinder or a partial sphere, as shown in Fig.~\ref{fig:shapes}.  In each case, we calculate the optimal curvature and determine how it depends on the asymmetry parameter $q$.  We then compare the energies of these two shapes to see which is favored, as a function of the film parameters.

\section{\label{sec:cylinder}Cylinder}

We first use Eq.~(\ref{eq:f2}) to calculate the energy when the film deforms into a partial cylinder, as shown in Fig.~\ref{fig:shapes}(a). For this calculation, we use an orthogonal curvilinear coordinate system $(\xi_1,\xi_2)$ on the midplane in the material frame, where $\xi_1$ is the direction around the circumference of the cylinder and $\xi_2$ is the direction along the axis. The 3D position of any point on the midplane can be written as
\begin{equation}
\label{eq:cRmid}
\mathbf{R}_{\mathrm{mid}}=r_c \cos \left( \frac{C_1 \xi _1}{r_c}\right)\hat{\mathbf{x}}+C_2  \xi _2\hat{\mathbf{y}}+r_c \sin \left(\frac{C_1 \xi _1}{r_c}\right)\hat{ \mathbf{z}},
\end{equation}
where $r_c$ is the cylinder radius, $C_1$ and $C_2$ are parameters that measure how much the coordinates elongate or shrink, and $\hat{\mathbf{x}}$, $\hat{\mathbf{y}}$, and $\hat{\mathbf{z}}$ are unit vectors in Cartesian coordinates for the lab frame. From Eq.~(\ref{eq:tmetric}), the target metric of the midplane is
\begin{equation}
\label{eq:tmetriccylinder}
\bar{a}_{\alpha\beta}=\left( \begin{array}{cc} p^2 & 0 \\ 0 & p^2\end{array} \right).
\end{equation}
From Eq.~\ref{eq:cRmid}, the actual metric and curvature tensors (first and second fundamental forms) of the partial cylinder are given by
\begin{equation}
a_{\alpha\beta}=\left( \begin{array}{cc} C_1^2 & 0 \\ 0 & C_2 ^2 \end{array} \right)
\end{equation}
and
\begin{equation}
b_{\alpha\beta}=\left( \begin{array}{cc} -C_1^2/r_c & 0 \\ 0 & 0 \end{array} \right).
\end{equation}
The elasticity tensor $\mathring{\mathcal{A}}^{\alpha\beta\gamma\delta}$ can be calculated by noticing that the target metric tensor before swelling is just the identity matrix, independent of position. Hence, the non-zero terms in the elasticity tensor are $\mathring{\mathcal{A}}^{1111}=\mathring{\mathcal{A}}^{2222}=4\mu$ and $\mathring{\mathcal{A}}^{1122}=\mathring{\mathcal{A}}^{1212}= \mathring{\mathcal{A}}^{2121}=\mathring{\mathcal{A}}^{2211}=2\mu$.

We now insert the target metric, actual metric, and curvature tensors into the 2D elastic energy of Eq.~(\ref{eq:f2}), and minimize over the parameters $C_1$, $C_2$, and $r_c$.  To lowest order in $w$, we find that the stretching factors are $C_1=C_2=p$, the cylinder radius is
\begin{equation}
\label{eq:rc}
r_c=\frac{2p^2}{3q},
\end{equation}
and the corresponding 2D elastic energy density at that minimum is
\begin{equation}
\label{eq:f2cylinder}
f_2=\frac{29q^2}{8p^2}w^3 \mu.
\end{equation}
These results imply that the curvature tensor $b_{\alpha\beta}$ is proportional to $q$, and that the strain tensor $\varepsilon_{\alpha\beta}^{\mathrm{2D}}=0$; i.~e.\ there is no strain \emph{on the midplane}.  Note, however, that there is still nonzero shear strain \emph{off the midplane} because the metric tensor off midplane shows anisotropic swelling while the target metric favors isotropic swelling; that is the reason why the elastic energy is non-zero.  In the limit of a symmetric film with $q\to 0$, then $r_c\to\infty$, the film remains flat, and there is only uniform swelling without energy cost.

Equation~(\ref{eq:f2cylinder}) shows that the elastic energy density of the partial cylinder is uniform in 2D, independent of position on the midplane, and hence the total elastic energy is just proportional to the film area.  In particular, if the initial shape of the film is a disk with radius $r_{\mathrm{max}}$ in the material frame, hence radius $p r_{\mathrm{max}}$ in the midplane after swelling, then the total elastic energy is
\begin{equation}
\label{eq:Fc}
F_c=\frac{29}{8}\pi q^2 r_{\mathrm{max}}^2 w^3 \mu.
\end{equation}

\section{\label{sec:sphere}Sphere}

Let us now consider a circular thin film deforming into a partial of a sphere, as shown in Fig.~\ref{fig:shapes}(b).  For this problem, it is convenient to use polar coordinates $(r,\phi)$ on the midplane in the material frame, where $r$ is the radial displacement from the center of the circular film before swelling, and $\phi$ is the azimuthal angle, which is assumed not to change during swelling and deformation.  The 3D position of any point on the midplane can then be written as
\begin{equation}
\label{eq:sRmid}
\mathbf{R}_{\mathrm{mid}}=r_s\sin\theta(r)\cos\phi\hat{\mathbf{x}}
+r_s\sin\theta(r)\sin\phi\hat{\mathbf{y}}+r_s\cos\theta(r)\hat{\mathbf{z}},
\end{equation}
where $r_s$ is the radius of the sphere and $\theta(r)$ is a monotonically increasing function of $r$, to be determined, which describes how the material stretches or shrinks in the radial direction.  (In particular, $\theta(r)$ gives the angular position on the partial sphere up from the $(-z)$-axis corresponding to the radial position $r$ in the material frame.) In this coordinate system, the target metric of the middle plane is
\begin{equation}
\bar{a}_{\alpha\beta}=\left(\begin{array}{cc} p^2 & 0 \\ 0 & p^2r^2 \end{array}\right).
\end{equation}
Note that this target metric is equivalent to Eq.~(\ref{eq:tmetriccylinder}), but in a different coordinate system.  From Eq.~(\ref{eq:sRmid}), the actual metric and curvature tensors (first and second fundamental forms) are given by
\begin{equation}
a_{\alpha\beta}=\left(\begin{array}{cc} r_s^2 \theta '(r)^2 & 0 \\ 0 & r_s^2 \sin ^2 \theta(r) \end{array}\right),
\end{equation}
and
\begin{equation}
b_{\alpha\beta}=\left(\begin{array}{cc} -r_s \theta '(r)^2 & 0 \\ 0 & -r_s \sin ^2 \theta(r) \end{array}\right).
\end{equation}
In this coordinate system, the non-zero components of the elasticity tensor $\mathring{\mathcal{A}}^{\alpha\beta\gamma\delta}$ are $\mathring{\mathcal{A}}^{1111}=4\mu$, $\mathring{\mathcal{A}}^{2222}=4\mu/r^4$, and $\mathring{\mathcal{A}}^{1122}=\mathring{\mathcal{A}}^{1212}= \mathring{\mathcal{A}}^{2121}=\mathring{\mathcal{A}}^{2211}=2\mu/r^2$.

The 2D elastic energy of Eq.~(\ref{eq:f2}) is now a functional of the stretching function $\theta(r)$ and the sphere radius $r_s$.  To minimize this energy over $\theta(r)$, we expand $\theta(r)$ as a power series in $r$ and minimize over the series coefficients.  The leading terms are then
\begin{equation}
\theta(r)=\pi-(p-\frac{p^3w^2}{6r_s^2}+\frac{3pqw^2}{8r_s})\frac{r}{r_s}.
\end{equation}
To minimize the energy over $r_s$, we must consider two regimes in terms of $q$, $w$, and $r_{\mathrm{max}}$, the radius of the film in the material frame.  In the first regime, where $q r_{\mathrm{max}}^2/w\ll 1$, the optimum sphere radius is $r_s = p^2/q$, and the total elastic energy is
\begin{equation}
\label{eq:Fs1}
F_{\mathrm{s}}=\frac{7}{2}\pi q^2 r_{\mathrm{max}}^2 w^3\mu.
\end{equation}
By comparison, in the second regime where $q r_{\mathrm{max}}^2/w \gg 1$, the sphere radius is $r_s = (pr_{\mathrm{max}})^{4/3} (26qw^2)^{-1/3}$, and the total elastic energy is
\begin{equation}
\label{eq:Fs2}
F_{\mathrm{s}}=4\pi q^2 r_{\mathrm{max}}^2 w^3 \mu.
\end{equation}

\section{\label{sec:discuss}Discussion}

Our calculations in the preceding sections lead to specific conclusions about the cylindrical and spherical shapes, as well as more general insight into elastic theory for asymmetric films.

For the specific problem of cylindrical and spherical shapes, we can see that Eq.~(\ref{eq:Fc}) for the elastic energy of a partial cylinder is between the two regimes of Eqs.~(\ref{eq:Fs1}) and~(\ref{eq:Fs2}) for a partial sphere.  This result implies that the spherical deformation is favored for disks of small radius $r_{\mathrm{max}}<r_{\mathrm{critical}}$, while the cylindrical deformation is favored for disks of large $r_{\mathrm{max}}>r_{\mathrm{critical}}$.  Calculating the actual value of $r_{\mathrm{critical}}$ requires higher-order terms than we have presented here, and the numerical result is $r_{\mathrm{critical}}=1.6(wp/q)^{1/2}$.  To understand this result, note that the partial sphere has an isotropic deformation in the midplane, which is consistent with the target metric, while the partial cylinder breaks is anisotropic in the midplane and disagrees with the target, which costs extra energy.  For that reason, the partial sphere is favored for small $r_{\mathrm{max}}$.  By contrast, for large $r_{\mathrm{max}}$ the partial sphere must have extra stretching in the midplane, and hence it becomes disfavored with respect to the partial cylinder.

More generally, we have shown that the theoretical approach of Sharon and collaborators can be applied to asymmetric films.  Through this approach, we transform the 3D elastic energy into the effective 2D elastic energy of Eq.~(\ref{eq:f2}).  This effective 2D elastic energy shows the standard stretching and bending energies, which have been studied extensively for symmetric films, as well as new terms arising from the asymmetry.  These new terms include a spontaneous curvature term, which is linear in the curvature tensor and hence favors curvature of the asymmetric film, as well as a coupling between spontaneous curvature and in-plane strain.  These new terms should provide new opportunities to design synthetic materials that will spontaneously form desired shapes for technological applications.

\acknowledgments

This work was supported by the National Science Foundation through Grants DMR-0605889 and 1106014.

\bibliography{asymmetric_film}

\end{document}